\documentclass[twocolumn]{aastex63}
\usepackage{natbib,amsmath,scalerel,placeins}

\usepackage{graphicx}
\usepackage{txfonts}
\usepackage{xspace}
\usepackage{color}
\usepackage{url}
\usepackage{hyperref}
\usepackage{comment}
\usepackage[flushleft]{threeparttable}

\newcommand{\lum}{erg s$^{-1}$}
\newcommand{\Nu}{{\it NuSTAR\xspace}}
\newcommand{\ch}{{\it Chandra\xspace}}
\newcommand{\xm}{{\it XMM-Newton\xspace}}

\newcommand{\src}{XTE\,J1858+034\xspace}

\newcommand{\rmn}[1] {{\rm #1}}

\def\arcdeg{\hbox{$^\circ$}}
\def\arcmin{\hbox{$^\prime$}}
\def\arcsec{\hbox{$^{\prime\prime}$}}

\def\flux{erg s$^{-1}$ cm$^{-2}$}

\def\lum{erg s$^{-1}$}

\shorttitle{Properties of XRP XTE J1858+034}
\shortauthors{Tsygankov et al.}
\begin{document}

\title{X-ray pulsar XTE\,J1858+034: discovery of the cyclotron line and the revised optical identification}

\author[0000-0002-9679-0793]{Sergey~S.~Tsygankov}
\affiliation{Department of Physics and Astronomy, FI-20014 University of Turku,  Finland}
\affiliation{Space Research Institute of the Russian Academy of Sciences, Profsoyuznaya Str. 84/32, Moscow 117997, Russia}
\correspondingauthor{Sergey~S.~Tsygankov}
\email{sergey.tsygankov@utu.fi}

\author[0000-0002-6255-9972]{Alexander~A.~Lutovinov}
\affiliation{Space Research Institute of the Russian Academy of Sciences, Profsoyuznaya Str. 84/32, Moscow 117997, Russia}

\author[0000-0002-5983-5788]{Sergey~V.~Molkov}
\affiliation{Space Research Institute of the Russian Academy of Sciences, Profsoyuznaya Str. 84/32, Moscow 117997, Russia}

\author[0000-0001-6316-9880]{Anlaug~A.~Djupvik}
\affiliation{Nordic Optical Telescope, Apartado 474, 38700 Santa Cruz de La Palma, Santa Cruz de Tenerife, Spain}
\affiliation{Department of Physics and Astronomy, Aarhus University, NyMunkegade 120, DK-8000 Aarhus C, Denmark}

\author{Dmitri~I.~Karasev}
\affiliation{Space Research Institute of the Russian Academy of Sciences, Profsoyuznaya Str. 84/32, Moscow 117997, Russia}

\author[0000-0001-8162-1105]{Victor~Doroshenko}
\affiliation{Institut f\"ur Astronomie und Astrophysik, Universit\"at T\"ubingen, Sand 1, D-72076 T\"ubingen, Germany}
\affiliation{Space Research Institute of the Russian Academy of Sciences, Profsoyuznaya Str. 84/32, Moscow 117997, Russia}

\author[0000-0003-2306-419X]{Alexander~A.~Mushtukov}
\affiliation{Leiden Observatory, Leiden University, NL-2300RA Leiden, The Netherlands}
\affiliation{Space Research Institute of the Russian Academy of Sciences, Profsoyuznaya Str. 84/32, Moscow 117997, Russia}
\affiliation{Pulkovo Observatory, Russian Academy of Sciences, Saint Petersburg 196140, Russia}

\author[0000-0002-0380-0041]{Christian~Malacaria}
\affiliation{NASA Marshall Space Flight Center, NSSTC, 320 Sparkman Drive, Huntsville, AL 35805, USA}
\affiliation{Universities Space Research Association, Science and Technology Institute, 320 Sparkman Drive, Huntsville, AL 35805, USA}

\author[0000-0001-9840-2048]{Peter~Kretschmar}
\affiliation{European Space Agency (ESA), European Space Astronomy Centre (ESAC), Camino Bajo del Castillo s/n, 28692 Villanueva de la Cañada, Madrid, Spain}

\author[0000-0002-0983-0049]{Juri~Poutanen}
\affiliation{Department of Physics and Astronomy, FI-20014 University of Turku, Finland}
\affiliation{Space Research Institute of the Russian Academy of Sciences, Profsoyuznaya Str. 84/32, Moscow 117997, Russia}
\affiliation{Nordita, KTH Royal Institute of Technology and Stockholm University, Roslagstullsbacken 23, SE-10691 Stockholm, Sweden}

\begin{abstract}
We present results of a detailed investigation of the poorly studied X-ray pulsar \src\ based on the data obtained with the \Nu\ observatory during the outburst of the source in 2019. The spectral analysis resulted in the discovery of a cyclotron absorption feature in the source spectrum at $\sim$48~keV both in the pulse phase averaged and resolved spectra. Accurate X-ray localization of the source using the \Nu\ and {\it Chandra} observatories allowed us to accurately determine the position of the X-ray source and identify the optical companion of the pulsar. The analysis of the counterpart properties suggested that the system is likely a symbiotic binary hosting an X-ray pulsar and a late type companion star of K-M classes rather than Be X-ray binary as previously suggested.
\end{abstract} 


\section{Introduction}
The transient X-ray pulsar (XRP) \src was discovered with the All Sky Monitor onboard the {\it RXTE} observatory in 1998 \citep{1998IAUC.6826....2R}. Pulsations with a period of $221.0\pm0.5$~s detected during the same outburst in the {\it RXTE}/PCA data \citep{1998IAUC.6826S...1T} and the transient nature of \src led these authors to conclude that the system is likely an XRP with Be counterpart.
A long-term light curve based on the {\it RXTE}/ASM data clearly shows regular outbursts with a period of $\sim$380\,d, which was interpreted as the orbital period of the binary system \citep{2008ARep...52..138D}. 

The energy spectrum of \src is typical for XRPs and can be described as an absorbed cut-off power law with an equivalent hydrogen column density of $6\times10^{22}$~cm$^{-2}$, modified by an iron emission line around 6.6\,keV \citep{1998A&A...337..815P}. A similar spectral shape and spectral parameters were obtained from {\it INTEGRAL} data \citep{2005AstL...31..729F,2008ARep...52..138D} collected during an outburst in 2004 \citep{2004ATel..274....1M}. Evidence of a cyclotron absorption line has never been reported for \src\ in the literature, therefore, the magnetic field strength of the neutron star (NS) remains uncertain. On the other hand, \cite{1998A&A...337..815P} discovered quasi-periodic oscillations at 0.11\,Hz, which is significantly higher than the spin frequency. Interpreting this feature in the frame of the beat frequency model, they estimated the NS magnetic field to be $0.8\times10^{12} r_{\rm kpc}$ G, where $r_{\rm kpc}$ is the distance to the system in kpc.

The X-ray localisation of \src was originally obtained from the {\it RXTE} data \citep{1998IAUC.6828....2M} and later improved by \cite{2004ATel..274....1M} using the observations with {\it INTEGRAL}/JEM-X 
and {\it INTEGRAL}/IBIS/ISGRI.
Based on these results \cite{2005A&A...440..637R} proposed a star with the coordinates 
RA=$18^{\rmn{h}}58^{\rmn{m}}36^{\rmn{s}}$, Dec=$3^\circ26\arcmin09\arcsec$,
as a possible counterpart for the source. Optical spectroscopy indicated that it was the only nearby star exhibiting H$\alpha$ emission, however, its position was not consistent with the JEM-X uncertainty. Therefore, this association can only be considered tentative.

The relative faintness of \src\ even during its outbursts and shortage of available data did not allow to make any definitive conclusions regarding the physical properties of the NS in the system up to date. Here we report the results of a {\it NuSTAR} observation of \src\ performed during its outburst in the Fall 2019 \citep[detected by the MAXI instrument,][]{2019ATel13217....1N}, which allowed to conduct detailed spectral and timing analysis and localize the source. The localization of the source was then further refined using an archival {\it Chandra} observation, and subsequent near-IR identification of the companion star using the Nordic Optical Telescope (NOT) telescope.

\section{Observations and data reduction}

This work is based on several data sets in X-ray (\ch, \xm\ and \Nu\ observatories) and near-IR (the UKIDSS survey and NOT telescope) bands. The complete list of the data used is presented in Table~\ref{tab:data}.

\begin{table}
    \centering
    \caption{List of \src\ observations utilized in this work.}
    \begin{tabular}{cccc}
    \hline
    ObsID & Start date & Start MJD & Exposure (ks) \\
    \hline
    \multicolumn{4}{c}{\xm}  \\
    0302970201 & 2006-04-11 & 53836.91 & 25 \\
    \multicolumn{4}{c}{\ch}  \\
    14645 & 2013-02-17 & 56340.17 & 5  \\
    \multicolumn{4}{c}{\Nu}  \\
    90501348002 & 2019-11-03 & 58790.29 & 43 \\
    \multicolumn{4}{c}{Nordic Optical telescope}  \\
    STzd200003-7     & 2016-04-21 & 57499.11 & 0.12$^a$\\ 
    NCzd200587-640   & 2016-04-21 & 57499.11 & 0.54$^b$\\
    NCzi150446-449   & 2016-09-15 & 57646.99 & 2.4  \\
    \hline
    \end{tabular}
    	\begin{tablenotes}
        \item $^{a}$ Per filter $BVRi$. 
        \item $^{b}$ Per filter $JHK_s$.  
        \end{tablenotes}

    \label{tab:data}
\end{table}

\subsection{\Nu\ observatory}

The \Nu\ observatory includes two identical co-aligned X-ray telescopes focusing X-ray photons onto two Focal Plane Modules A and B (FPMA and FPMB) \citep{2013ApJ...770..103H}. In the context of the current work the most important \Nu\ capabilities are the moderately high imaging resolution of 18$\arcsec$ (full width at half maximum) and unprecedented sensitivity in hard X-rays from 3 to 79 keV. \src was observed with \Nu\ on November 3, 2019 with an exposure of $\sim$43 ks (see Table~\ref{tab:data})  near the peak of the outburst. 

The raw data were reduced following the standard procedures described in the {\it NuSTAR} user guide and using the standard {\it NuSTAR}
Data Analysis Software ({\sc nustardas}) v1.8.0  with CALDB version 20190513.
The source spectrum was extracted from a source-centered circular region with
radius of 80\arcsec\ using the {\sc nuproducts} routine. The background
was extracted from a source-free circular region with radius of 80\arcsec\ in
the corner of the field of view. In order to apply standard $\chi^2$ minimization procedures, the original source spectrum was rebinned in order to have at least 25 counts per energy bin. The following spectral analysis was performed using {\sc xspec} package \citep{1996ASPC..101...17A}.

\subsection{\xm\ observatory}

In order to investigate different states of \src, the \xm\ data obtained during the quiescent state of the source were also analysed. The source was observed in April 2006 for $\sim$25 ks. EPIC pn and EPIC MOS detectors were operated in the full frame mode with medium filters. The data reduction procedures using the \xm\  {\sc Science Analysis Software} (SAS; version 18.0) were applied following standard procedures.\footnote{\url{https://www.cosmos.esa.int/web/xmm-newton/sas-threads}} 
After the standard pipeline processing, we searched for possible intervals of high background and rejected them. This resulted in lowering of the effective exposure down to $\sim$16 ks and $\sim$6 ks for MOS and pn, respectively.
For the analysis of the EPIC data, we selected events with patterns in the range 0–4 for the pn camera and 0–12 for the two MOS cameras, using a circular region with a radius of 20\arcsec\ around the source positions. The circular background region with a radius of 30\arcsec\ was placed close to the source at the same CCD and avoiding other point sources. 

\subsection{\ch\ observatory}

\src was observed by \ch\ advanced CCD Imaging Spectrometer \citep[ACIS; ][]{2003SPIE.4851...28G} on February 17, 2013 with an exposure of $\sim5$~ks as part of the project aimed at studies of the transient XRPs in a quiescent state \citep{2017MNRAS.470..126T}.
For the data reduction we used the software package {\sc ciao} v4.12 with an up-to-date {\sc caldb} v4.9.1. The procedure {\sc celldetect} was used to determine coordinates of sources in the \ch\ field of view. The source spectrum was extracted from a circular aperture with a radius of 3$\arcsec$ around the position, while for the background extraction we used a circular region near the source with a radius of 15$\arcsec$.

\subsection{Nordic Optical Telescope}

Optical and near-IR observations were obtained at the Nordic Optical Telescope (NOT) through applications in the fast-track service \citep{2010ASSP...14..211D}, using the Standby CCD camera StanCam and the Nordic Optical Telescope near-IR Camera and spectrograph (NOTCam\footnote{For details about NOTCam, we refer to \url{http://www.not.iac.es/instruments/notcam/}.}). StanCam images in $BVRi$ bands, as well as NOTCam high spatial resolution images ($0\farcs079$/pix) in the $JHK_s$ bands were obtained on April 20, 2016 in good seeing conditions (FWHM = $0\farcs6$--$0\farcs7$). The near-IR images were obtained by small step dithering in a 3$\times$3 pattern with 30\,s exposures in each position obtained in ramp-sampling mode, reading out every 5\,s, giving a total of 540\,s per filter in the combined $J$, $H$ and $K_s$ images. The StanCam exposures of 120\,s showed no detection in any band, while the near-IR images detected a red counterpart at the location of the X-ray source. A $K$-band spectrum was obtained on Sept 15, 2016 under less favourable conditions. The setup used was the WF camera ($0\farcs234$/pix), Grism \#1 with a dispersion of 4.1 \AA\ per pixel, the 128 micron slit ($0\farcs6$ wide), and the $K$-band filter (\#208) used as an order sorter, which gives a resolution of 10.5 \AA\ and a resolving power of $\lambda$/$\Delta \lambda$ = 2100. The spectra were obtained in the ABBA dithering mode, exposing 600\,s per position and using the ramp-sampling mode to read out non-destructively 10 times every 60\,s. In-situ arc and halogen lamps were observed to take out the effect of fringing as much as possible and to account for flexure in the wavelength calibration. A telluric standard close to the target was observed immediately before the target.

The near-IR images were reduced using the NOTCam IRAF package to do bad-pixel correction, flat-fielding with differential twilight flats, sky-subtraction, and shifting and median combining of the individual images. The individual 2D $K$-band spectra were hot-pixel and zero-pixel corrected, flat-fielded and sky subtracted before 1D extraction using standard IRAF tasks. The individual 1D spectra were wavelength calibrated and thereafter combined to a final spectrum. This was divided by the telluric standard spectrum to correct for atmospheric features and afterwards multiplied by a black body continuum of the same spectral type as the telluric standard to correct the slope in the spectrum. Due to mediocre sky conditions the final spectrum had a poor S/N ratio and was therefore smoothed over 17 pixels, lowering the resolving power to R = 120.

\section{Results}
\begin{figure}
\centering
\includegraphics[width=\columnwidth]{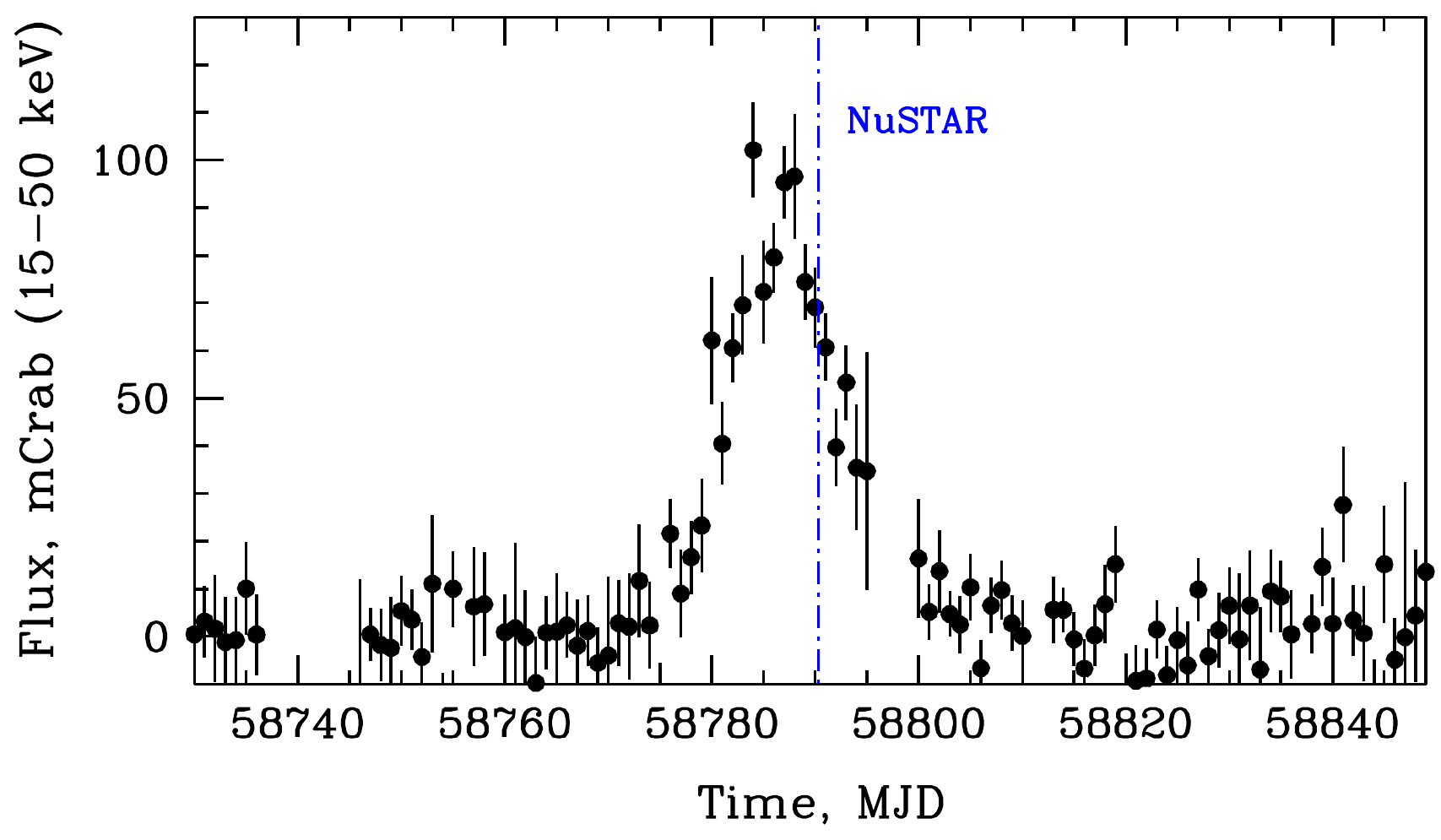}
\caption{Light curve of \src\ in 15-50 keV energy band obtained by the {\it Swift}/BAT monitor. The vertical dash-dotted line corresponds to the time of our {\it NuSTAR} observation.}\label{fig:lc}
\end{figure}

\src\ was observed by the {\it NuSTAR} observatory close to the peak of the 2019 outburst. The light curve of the source in the 15--50 keV energy band obtained\footnote{\url{http://swift.gsfc.nasa.gov/results/transients/}} by the Burst Alert Telescope on board the Neil Gehrels Swift Observatory \citep[\textit{Swift}/BAT; ][]{2013ApJS..209...14K} is shown in Fig.~\ref{fig:lc}. 
The relatively high brightness of the source allowed us to study its properties in a broad energy band in details. 
At the same time soft band \ch\ and \xm\ data were collected in the low state with low counting statistics which was insufficient to detect pulsations. All uncertainties in the paper are reported at 1$\sigma$ confidence level, unless otherwise stated.

\subsection{Pulse profile and pulsed fraction}
\label{sec:pprof}

Orbital parameters except for the orbital period are not known for \src. Considering that the available X-ray data only covers a small fraction of the orbit, it was also impossible to determine those based on the X-ray timing.
Therefore, for the timing analysis only the barycentric but no binary correction was applied to the light curves.  Using standard epoch folding technique \citep[{\sc efsearch} task of {\sc ftools}; ][]{1987A&A...180..275L} strong pulsations in the full band were found with the period $P=218.382(2)$~s. The uncertainty for the pulse period value was determined from the simulated light curves following the procedure described by \citet{2013AstL...39..375B}. 

\begin{figure}
\centering
\includegraphics[width=\columnwidth]{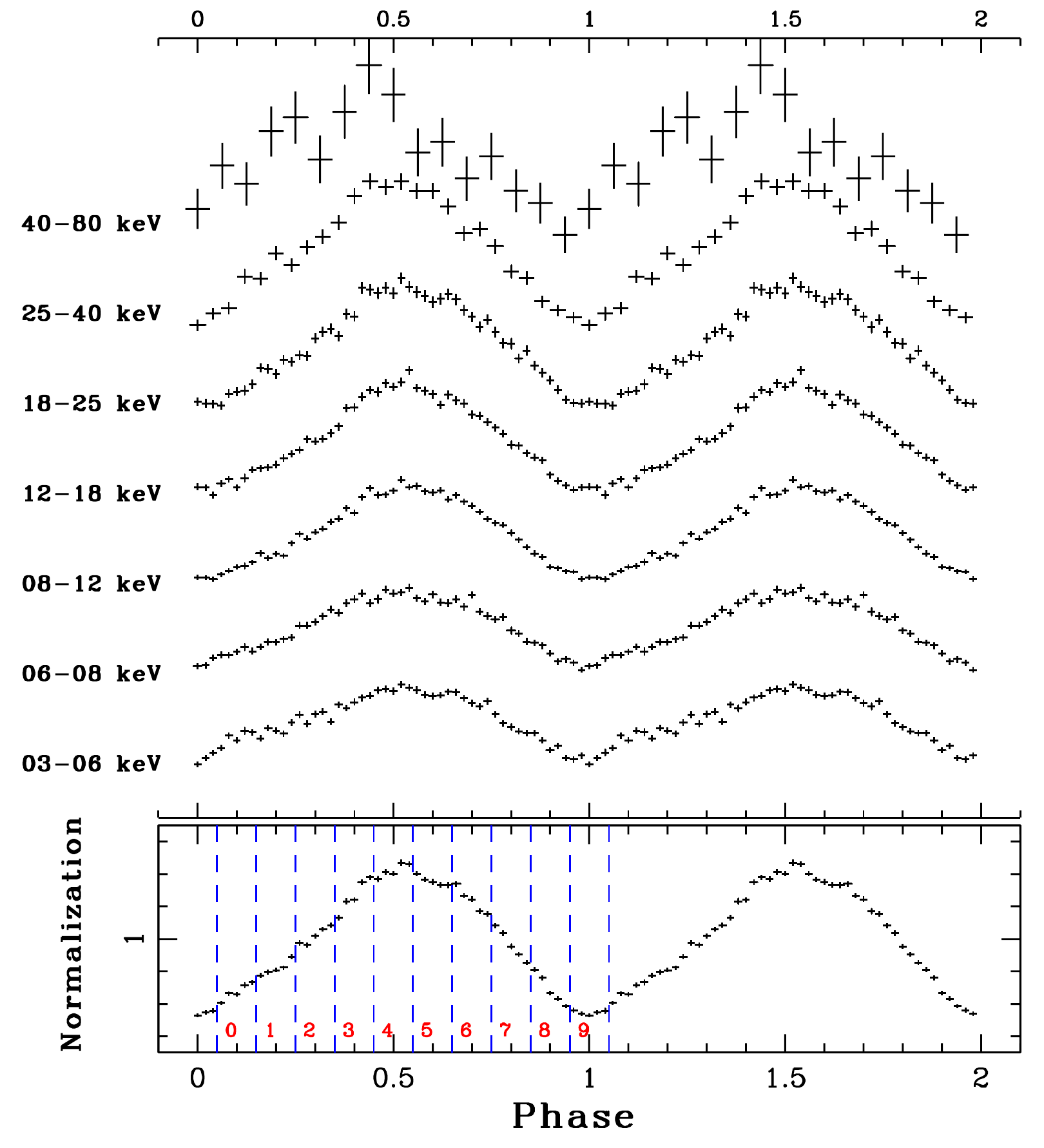}
\caption{{\it Top:} Pulse profiles of \src\ obtained with {\it NuSTAR} in
  different energy bands. The profiles are shown twice in phase and arbitrarily rescaled in count rate for clarity.
  {\it Bottom:} Pulse profile of \src in the full energy band 3--79 keV with phase bins selected for the phase-resolved spectral analysis shown with vertical dashed lines.}\label{fig:pprof}
\end{figure}

High count statistics allowed us to reconstruct the pulse profile of the source in several energy bands (selected to provide sufficient number of photons in each) from 3 to 79 keV (see Fig.~\ref{fig:pprof}). 
Even at the highest energies (in the 40--80 keV band) where counting statistics is limited, pulsations at an expected period are significantly detected in a blind search (Lomb-Scargle false alarm probability $p\sim2.8\times10^{-7}$ for over 2.5 millions of trial periods).
The overall shape of the pulse profile is sine-like single-peaked, consistent with the results from the {\it INTEGRAL} observatory obtained by \citet{2008ARep...52..138D}. We note, however, that although the pulse profile shape remains constant at different energies, some tentative sign of a phase lag is observed with the soft profile lagging the hard one.

\begin{figure}
\centering
\includegraphics[width=\columnwidth]{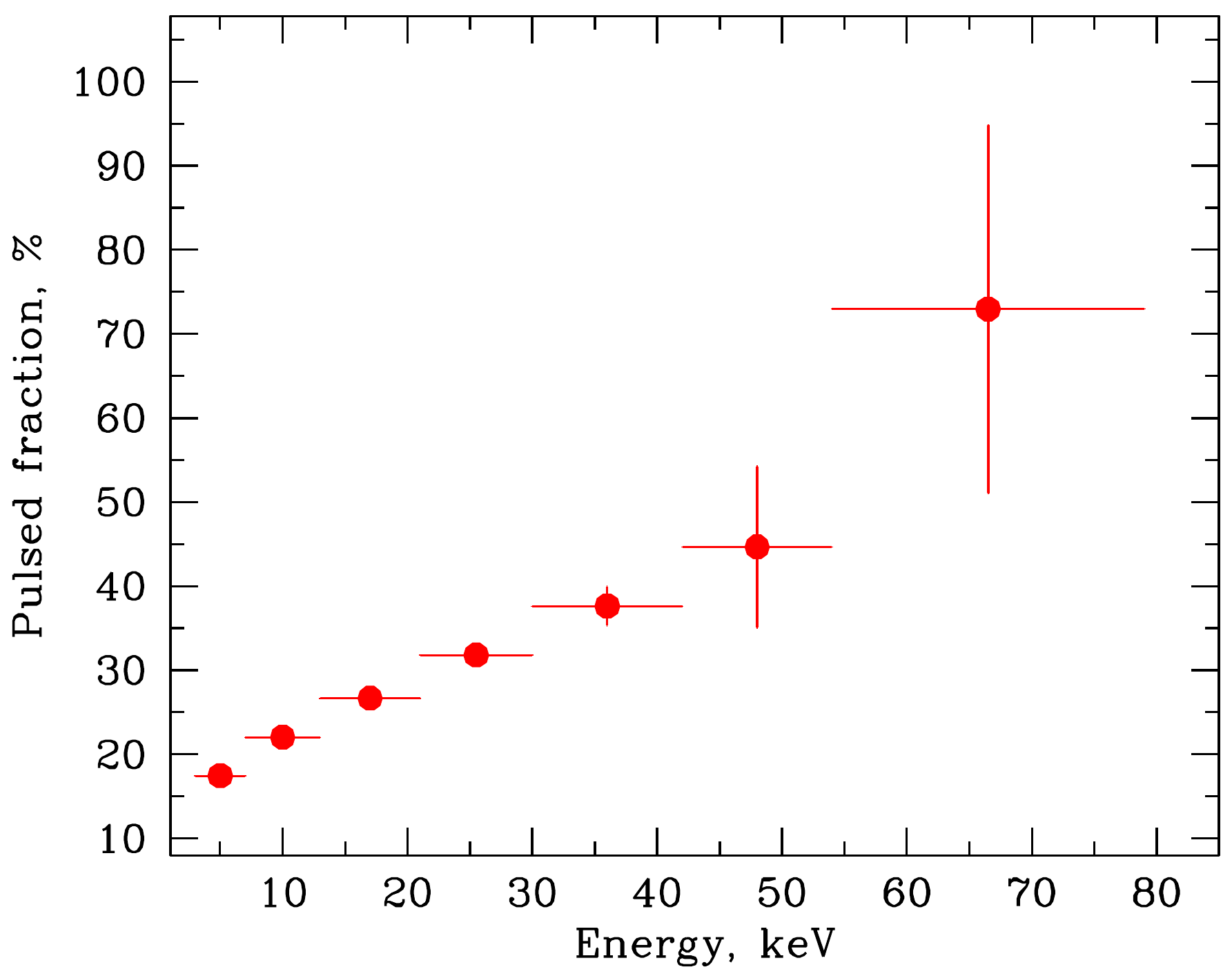}
\caption{Dependence of the pulsed fraction of \src\ on energy based on the {\it NuSTAR} data.}\label{fig:pfrac}
\end{figure}

Based on the energy-resolved pulse profiles with 12 phase bins, we also calculated the pulsed
fraction\footnote{$\mathrm{PF}=(F_\mathrm{max}-F_\mathrm{min})/(F_\mathrm{max}+F
_\mathrm{min})$, where $F_\mathrm{max}$ and $F_\mathrm{min}$ are maximum and
minimum fluxes in the pulse profile, respectively.} as function of energy presented in Fig.~\ref{fig:pfrac}. 
The linear increase of the pulsed
fraction towards the higher energies can be clearly seen, which is consistent with typical behavior of the most of XRPs \citep{2009AstL...35..433L}. 

\subsection{Phase-averaged spectral analysis}
\label{phaver}

Previously the spectral properties of \src in X-ray band were studied using  {\it RXTE} and {\it INTEGRAL} data \citep{1998A&A...337..815P,2005AstL...31..729F,2008ARep...52..138D}. In both cases it was concluded that the source spectrum can be fit with an absorbed power law with a high energy exponential cut-off, i.e. the typical spectrum for XRPs. No evidence for other features such as  cyclotron lines was reported.
The {\it NuSTAR} observatory, owing to a sufficiently high energy resolution and much better sensitivity at high energies, allowed us to conduct a much more detailed search for the possible cyclotron lines in the source spectrum.

\begin{figure}
\centering
\includegraphics[width=\columnwidth]{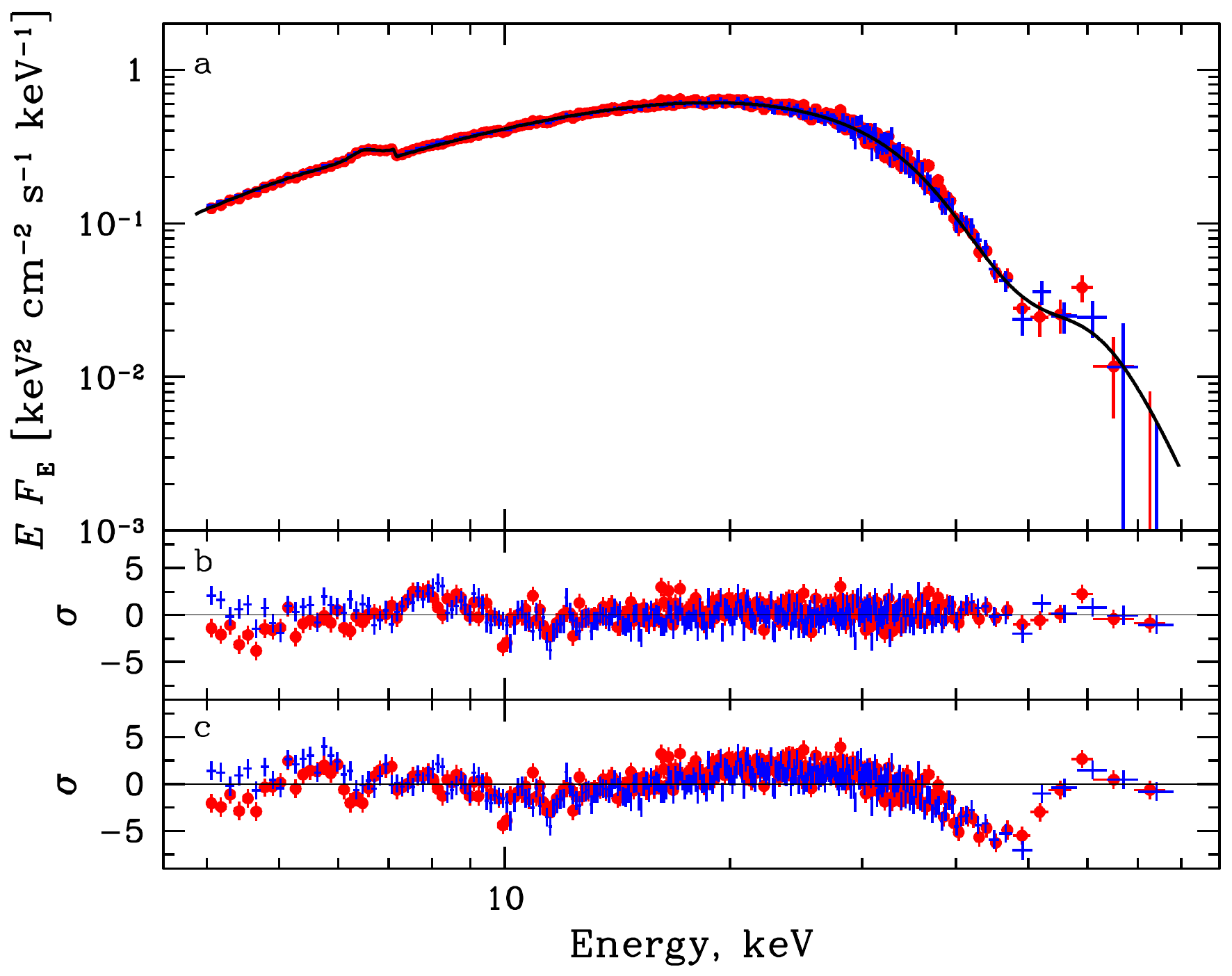}
\caption{(a) Unfolded X-ray spectrum of \src\ obtained with the \Nu\ observatory. The data from the FPMA and FPMB modules are shown with red and blue colors, respectively. Solid black line represents the best-fit model. 
(b) Residuals from the best-fit model for the \Nu\ data. (c) Residuals from the best-fit model not including the cyclotron absorption line.
}\label{fig:spe}
\end{figure}

\begin{table}
  \noindent
\caption{Best-fit parameters for a {\sc phabs$\times$(gau+comptt)$\times$gabs} model obtained for the \Nu\ observation. }
\label{tab:spe}
\begin{tabular}{ccc}
\hline
\hline
Parameter &     Units                   &           Value \\
\hline
$N_{\rm H}$ & $10^{22}\text{ cm}^{-2}$ &  $7.5\pm0.3$ \\
$E_{\rm Fe}$ &     keV                 &         $6.47\pm0.02$\\
$\sigma_{\rm Fe}$ &     keV             &           $0.22\pm0.03$\\
$N_{\rm Fe}$ &    $10^{-4}$ ph cm$^{-2}$ s$^{-1}$ &     $4.4\pm0.5$\\
$T_{\rm 0}^{a}$ &          keV           &  1.0$^{a}$  \\
$T^{b}$         &          keV          &  $5.63\pm0.01$ \\
$\tau^{c}$        &      --              &  $6.58\pm0.01$ \\
$E_{\rm cyc}$ &             keV        & $47.7\pm0.2$ \\
$\sigma_{\rm cyc}$ &         keV        & $8.15\pm0.13$ \\
Strength$_{\rm cyc}$ &      --              &  $20.6\pm0.5$ \\
$C_{\rm FPMA/FPMB}^d$  &     --        & $1.013\pm0.002$ \\
Flux$^e$ & $10^{-9}$\flux & $1.7_{-0.6}^{+0.4}$\\
$\chi^2$ (d.o.f.) &     --          &   1909.3 (1682)   \\
\hline
\end{tabular}
	\begin{tablenotes}
        \item $^{a}$ Input soft photons temperature. 3$\sigma$ upper limit. 
        \item $^{b}$ Plasma temperature. 
        \item $^{c}$ Plasma optical depth. 
        \item $^{d}$ Cross-normalisation factor between two \Nu\ modules. 
        \item $^{e}$ Unabsorbed flux in the 3--100 keV energy band. 
        \end{tablenotes}
\end{table}

Similarly to {\it RXTE} and {\it INTEGRAL}, the {\it NuSTAR} spectrum below $\sim$40~keV can be well described with several continuum models. In particular, we attempted to fit the data with several commonly used  for XRPs phenomenological continuum models, i.e. Comptonization model ({\sc comptt} in {\sc xspec}) and power law with high-energy exponential cutoff ({\sc cutoffpl} or {\sc po$\times$highecut} in {\sc xspec}). A more detailed study of different phenomenological and physical models applied to the source spectrum is presented in the accompanying paper by Malacaria et al. (in press).

To get an adequate fit, a photoelectric absorption at low energies ({\sc phabs} in {\sc xspec} assuming standard solar abundance from \citealt{1989GeCoA..53..197A}) and a fluorescent iron emission line ({\sc gau} in {\sc xspec}) were introduced to the model. 
However, irrespective of the continuum model we used, also residuals around $\sim45$\,keV in absorption are immediately apparent in the phase-averaged spectrum of \src. 
Figure~\ref{fig:spe} demonstrates the case of the {\sc comptt} continuum model (see panel c), but similar residuals also appear with other models. The fit can be greatly improved by inclusion of a gaussian absorption line in the model (Fig.~\ref{fig:spe}b).
The best-fit spectral parameters are presented in Table~\ref{tab:spe}. Inclusion of an absorption line with energy 47.7(2)\,keV  improves the fit from $\chi^2/{\rm dof}=2926.8/1685$ to $\chi^2/{\rm dof}=1909.3/1682$. The high statistical significance of the feature is obvious from the huge $\Delta\chi^2$ value. Its proper estimate using the {\sc xspec} script {\sc simftest} is technically infeasible due to an unrealistically large number of required simulations (the largest $\chi^2$ change obtained in $10^4$ simulations was only 48, which allows us to conclude that the significance estimated using this method must be high and in any case more than $3\sigma$).
Applying the {\it F}-test we calculated a false detection probability for the line $P=2.0\times10^{-155}$.

The detection of a cyclotron line at $\sim48$\,keV implies a magnetic field in \src\ of $\sim5.2\times10^{12}$\,G assuming a gravitational redshift $z=0.26$ for the typical NS parameters ($R=12$~km and $M=1.5 M_{\odot}$). 
Applying of the {\sc cyclabs} model instead of {\sc gabs} to describe the line results in a lower cyclotron energy of $\sim44$~keV. The discrepancy  between these two models is related to their definition and was found in other studies \citep[e.g.][]{2010ApJ...710.1755N,2015MNRAS.454.2714M,2017MNRAS.466.2143D}, see also the discussion of different line models in the review of \citet{2019A&A...622A..61S}.
We emphasize only that it should be kept in mind when comparing results from different studies.


In the quiescent state \src was observed twice: in April 2006 with \xm\ and in February 2013 with \ch. In both cases the source was found in the very low state with the flux about three orders of magnitude lower than in our \Nu\ observation. Both spectra were fitted with an absorbed power law ({\sc phabs$\times$pow} in {\sc xspec}) and black body ({\sc phabs$\times$bb} in {\sc xspec}) in order to determine the origin of the source emission in quiescence. A systematic study of the quiescent emission in transient XRPs with Be optical companions was conducted by \citet{2017MNRAS.470..126T}, however \src\ was excluded from their sample due to the uncertain nature of its optical counterpart.

Taking into account the small number of the collected photons, spectra in the low state were binned to have at least one count in each energy channel and fitted using the W-statistics \citep{1979ApJ...230..274W}. For the same reason it was impossible to constrain the absorption column simultaneously with other parameters. Therefore we fixed it at the best-fit value obtained from the \Nu\ data ($N_{\rm H}=7.5\times10^{22}$~cm$^{-2}$).
The best-fit spectral parameters in the soft X-ray band are presented in Table~\ref{tab:spelow}. It is clear that both models can fit the data equally well. 

We also fitted \xm/MOS and pn data jointly using the same simple models but with $N_{\rm H}$ as a free parameter in order to check if the absorption value depends on the luminosity state of the source. As a result for the black-body model we obtained $N_{\rm H}=(1.8^{+1.7}_{-1.1})\times10^{22}$~cm$^{-2}$ and temperature $kT=2.1^{+0.8}_{-0.5}$~keV, for the power law $N_{\rm H}=(3.8^{+2.6}_{-2.0})\times10^{22}$~cm$^{-2}$ and photon index $0.9^{+0.7}_{-0.6}$. With a similar quality of the fit we cannot make any final conclusion on the possibility of lower absorption in the quiescent state.

Both \ch\ and \xm\ observed \src shortly after the flares and measured  hard spectra (i.e. both have low photon indexes or high black-body temperatures) and fluxes. The hard spectral shape points to a likely non-thermal origin of the emission \citep{2017MNRAS.470..126T}. The source exhibits regular outburst activity and persistent accretion between flares cannot be excluded. Moreover, owing to the long spin period \src\ may belong to the group of pulsars accreting from a cold low-ionized disk even in quiescent state \citep{2017A&A...608A..17T}.

\begin{table}
  \noindent
\caption{Best-fit parameters for the {\sc phabs$\times$pow} and {\sc phabs$\times$bb} models
  obtained for both observations in low state. }
\label{tab:spelow}
\begin{tabular}{ccc}
\hline
\hline
Parameter &            \ch  &   \xm \\
\hline
\multicolumn{3}{c}{{\sc phabs$\times$pow} model} \\
\hline
$N_{\rm H}$ ($10^{22}\text{ cm}^{-2}$) & \multicolumn{2}{c}{7.5$^{\rm fix}$} \\
$\Gamma$         & $-0.4\pm0.7$ & $1.7\pm0.9$ \\
C-value (d.o.f.) &  27.5 (23)   & 105.5 (119) \\
Flux$^a$ (\flux) & $4.2_{-2.7}^{+0.2}\times10^{-13}$ & $4.3_{-3.1}^{+0.3}\times10^{-13}$ \\
\hline
\multicolumn{3}{c}{{\sc phabs$\times$bb} model} \\
\hline
$N_{\rm H}$ ($10^{22}\text{ cm}^{-2}$) & \multicolumn{2}{c}{7.5$^{\rm fix}$} \\
$kT_{\rm bb}$ (keV)  & $2.3_{-0.7}^{+1.8}$ & $1.4_{-0.4}^{+0.6}$ \\
C-value (d.o.f.) &  27.4 (23)  &  104.9 (119)\\
Flux$^a$ (\flux) & $3.5_{-1.6}^{+0.3}\times10^{-13}$ & $2.8_{-0.9}^{+0.4}\times10^{-13}$\\
\hline
\end{tabular}
	\begin{tablenotes}
        \item $^{a}$ Unabsorbed flux in the 0.5--10 keV energy band. 
        \end{tablenotes}
\end{table}

\subsection{Phase-resolved spectral analysis}
\label{phres}
\begin{figure}
\centering
\includegraphics[width=\columnwidth]{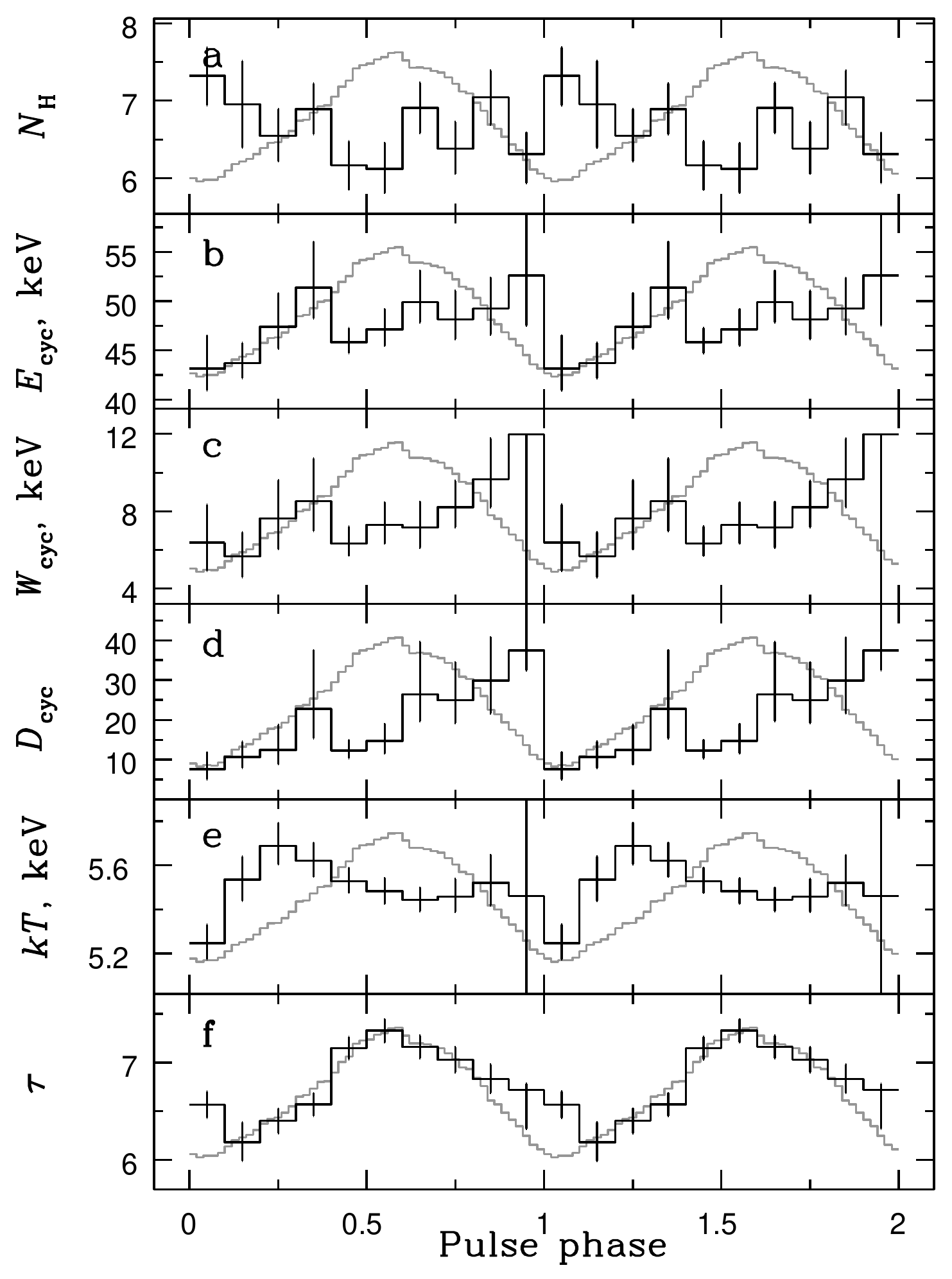}
\caption{Dependence of the spectral parameters of \src\ on the pulse phase revealed from the {\it NuSTAR} data. Different panels correspond to the absorption column in units of $10^{22}$~cm$^{-2}$ (a), cyclotron line energy (b), width (c) and strength (d), comptonizing plasma temperature (e) and optical depth (f). Pulse profile in 3-79 keV energy band is shown in grey for clarity.}\label{fig:speres}
\end{figure}

The high counting statistics of the \Nu\ data allowed us to perform a pulse phase-resolved spectral analysis. For that we used our best-fit model from the phase-averaged spectroscopy ({\sc phabs$\times$(gau+comptt)$\times$gabs}) and the phase-binning shown in Fig.~\ref{fig:pprof} which was defined based on the available counting statistics and observed pulse profile morphology. The source spectral parameters variations over the pulse are shown in Fig.~\ref{fig:speres}. 
Based on the {\sc simftest} simulations the significance of the cyclotron line in all phase-resolved spectra was shown to be higher than $3\sigma$.

We see that the continuum parameters (temperature and optical depth of the comptonizing plasma) vary significantly, whereas the relatively large uncertainty on the cyclotron line parameters prevent us from making any conclusion on their stability. The fitted values for absorption column, iron line parameters and the temperature of the seed photons stay constant within the errors over the pulse.

\subsection{X-ray position}

\begin{figure}
\begin{center} 
\includegraphics[width=\columnwidth]{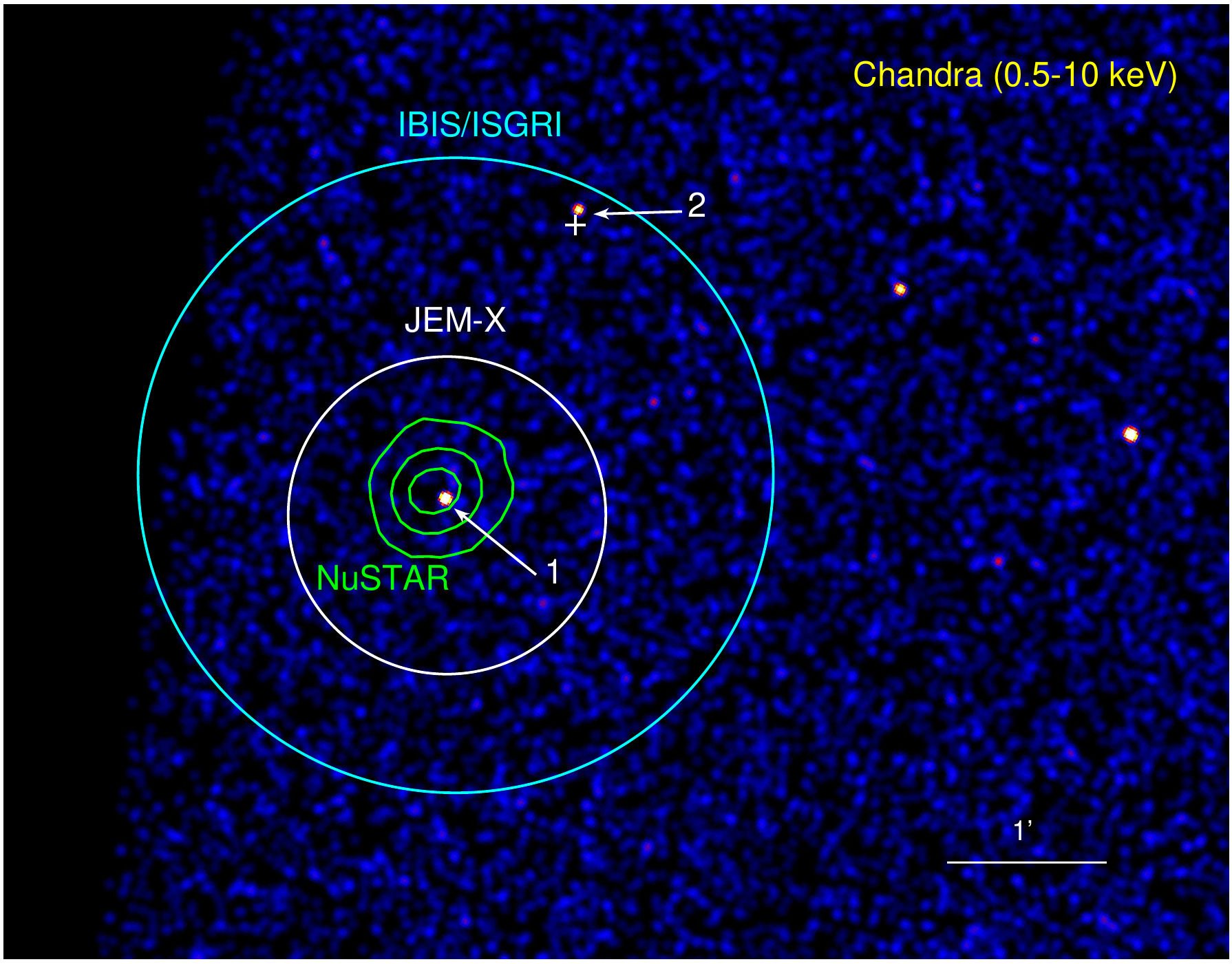}
\end{center}
\caption{Image of the sky field near \src\ obtained with the {\it Chandra} observatory in the 0.5-10 keV energy band. Localization uncertainties obtained from {\it INTEGRAL}/JEM-X and IBIS/ISGRI are shown with white and cyan circles, respectively. Green equal-flux contours represent the {\it NuSTAR} data analysed in this work. Numbers 1 and 2 correspond to the possible X-ray counterparts consistent with the IBIS/ISGRI localisation (see the text). A white cross marks the position of the optical counterpart proposed by \cite{2005A&A...440..637R}.}
\label{fig:chima} 
\end{figure}

The original localization accuracy of \src\ obtained with the {\it RXTE} observatory was 6\arcmin\ \citep{1998IAUC.6826....2R}. This was further improved by \cite{1998IAUC.6828....2M} based on repeated scans across the source with the {\it RXTE}/PCA instrument which allowed to localize the source at  
RA(2000)=$18^{\rmn{h}}58\fm6$, Dec(2000)=$3^\circ21\arcmin$,
with a 90-percent confidence error radius of 2\farcm5. Later, using the imaging capabilities of the JEM-X and IBIS telescopes onboard the {\it INTEGRAL} observatory, \cite{2004ATel..274....1M} further constrained coordinates of \src: 
RA=$18^{\rmn{h}}58^{\rmn{m}}43^{\rmn{s}}$, Dec=$3^\circ26\arcmin06\arcsec$,
for the IBIS/ISGRI data with the 2$\arcmin$ uncertainty, and 
RA=$18^{\rmn{h}}58^{\rmn{m}}44^{\rmn{s}}$,
Dec=$3^\circ26\arcmin02\arcsec$ for the JEM-X data with the 1$\arcmin$ uncertainty. The X-ray image of the sky field obtained with the {\it Chandra} observatory is shown in Fig.~\ref{fig:chima}. Two weak X-ray sources compatible with these localization regions were found in the data.

In order to determine the nature of \src, \cite{2005A&A...440..637R} performed optical photometric and spectroscopic observations of the field around the best-fit {\it INTEGRAL} position. It was revealed that only one star with the coordinates 
RA=$18^{\rmn{h}}58^{\rmn{m}}36^{\rmn{s}}$, Dec=$3^\circ26\arcmin09\arcsec$ 
exhibits H$\alpha$ emission (marked with a white cross in Fig.~\ref{fig:chima}). This star was proposed to be a possible counterpart of \src, since the counterpart was expected to be a Be-star. 

However, Fig.~\ref{fig:chima} demonstrates that this star cannot be an optical companion of \src\ (source marked with ``1''). This conclusion is confirmed by at least three facts: a coincidence of the localization regions obtained by {\it INTEGRAL}/JEM-X and \Nu\ (see green contours in Fig.~\ref{fig:chima}), detection of X-ray pulsations with the period of $\sim$221\,s by \Nu, and coincidence of \Nu\ and {\it Chandra} positions.  

\begin{figure}
\begin{center} 
\includegraphics[width=\columnwidth]{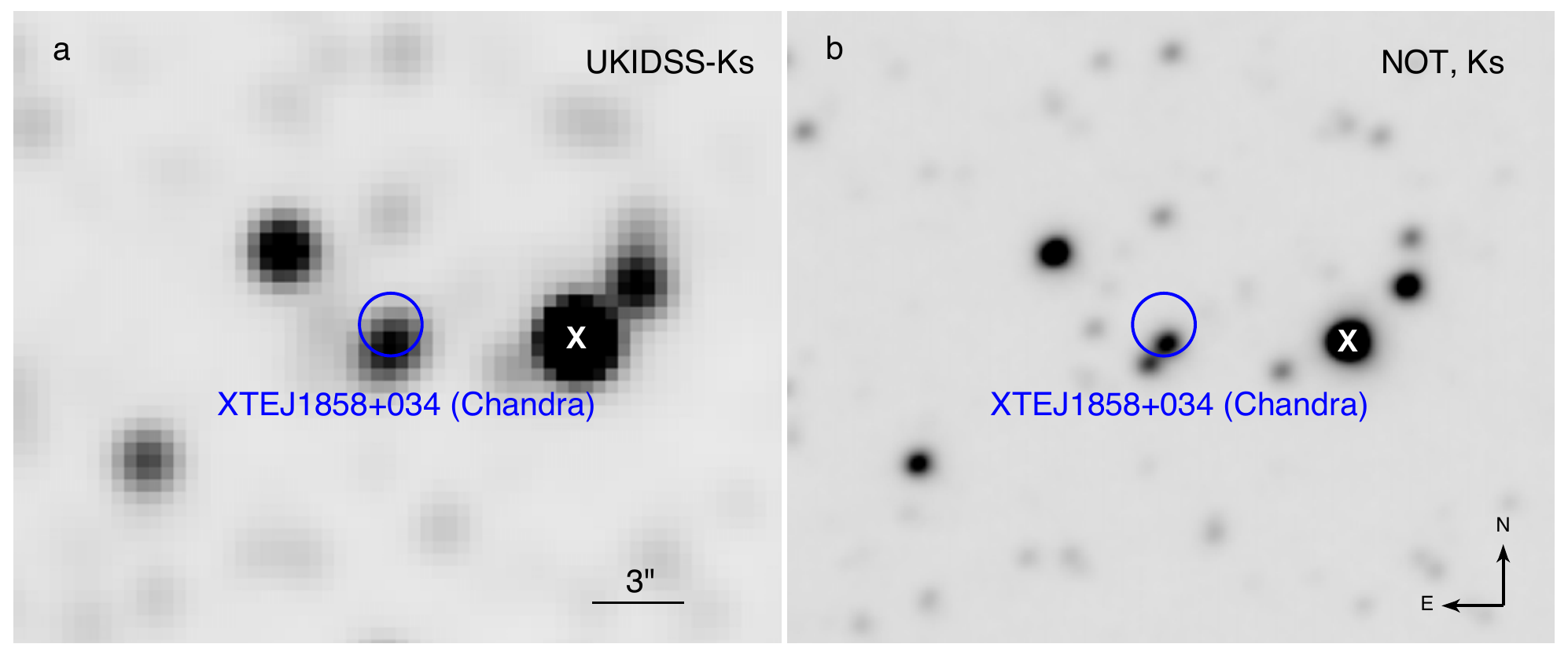}
\end{center}
\caption{Images of the sky containing \src\ obtained in the $K$-band from the GPS/UKIDSS sky survey data (left) and by the Nordic Optical Telescope (right).  The blue circle indicates the {\it Chandra} observatory error circle for the source position. The sign ''X'' indicates the star previously considered by \citet{2020ApJ...896...90M} as a possible companion.}
\label{fig:ukinot} 
\end{figure}

Using our {\it Chandra} data we obtained the precise coordinates for \src\ of RA=$18^{\rmn{h}}58^{\rmn{m}}43\fs64$, Dec=$3^\circ26\arcmin05\farcs8$ 
(J2000, marked with ``1'' in Fig.~\ref{fig:chima}),  
using the {\sc wavdetect} tool from the {\sc ciao} package. A statistical uncertainty of $0\farcs8$ at 90\% confidence level was obtained following the recommendations available on the online threads\footnote{\url{https://cxc.harvard.edu/ciao/threads/wavdetect/}}. Taking into account the systematic uncertainty of {\it Chandra} absolute positions of the same value\footnote{\url{https://cxc.harvard.edu/cal/ASPECT/celmon/}}, the resulting localization accuracy of the source we obtained is $1\farcs2$ (90\% confidence level radius; see blue circle in Fig.~\ref{fig:ukinot}).


\section{Discussion}

\subsection{Structure of the emitting region}

The observational properties of X-ray emission from XRPs depend on the physical conditions and geometrical structure of the emitting regions at the NS surface, and thus can be used to probe them.
As was already mentioned in Sect.~\ref{sec:pprof}, the pulse profile of \src\ has a sine-like single-peaked shape that is almost independent of energy. At the same time some tentative sign of a phase lag is observed with the soft profile lagging the hard one. A quantitative interpretation of this phenomenon is not possible in absence of adequate models describing emission of X-ray pulsars, however, qualitatively one may speculate that it may be associated with the resonant scattering of X-ray photons by the accretion flow above the hot spot. Indeed, considering that the source is likely in the sub-critical regime of accretion (see below), the optical thickness of the accretion flow above the hot spot is below unity for a non-resonant scattering. Thus, the low energy photons ($E<25$ keV) leave the system freely with a pencil-beam emission diagram \citep{1975A&A....42..311B}. However, the scattering cross-section near the cyclotron energy ($E_{\rm cyc}=48$ keV; see below) is well above unity \citep{1979PhRvD..19.2868H,1986ApJ...309..362D}. Because of that, the cyclotron photons are scattered by the accretion flow. Thus results in an energy-dependent beam function, leading to the observed lag between the pulse profiles.

Variations of the spectral continuum over the pulse phase are also consistent with a pencil-beam emission diagram. From Fig.~\ref{fig:speres} one can notice also that the optical depth $\tau$ appears to correlate with the flux in the profile, whereas the plasma temperature $kT$ shows an anti-correlation (except main minimum) reaching maximum values in the wings of the flux peak (see Fig.~\ref{fig:speres}e). This behavior can be interpreted in terms of the sub-critical accretion onto strongly magnetized NS with a pencil-beam emission diagram. In this case the accretion flow loses its kinetic energy in the atmosphere of a NS, resulting in the inverse temperature profile in the NS atmosphere with hotter upper layers where most of energy is released \citep{1975A&A....42..311B}. The maximal flux in the pulse profile corresponds to the situation when an observer looks at a hot spot close to the local normal. Then photons from the deeper and colder layers are detected resulting in a negative/positive correlation of temperature/optical depth with the photon energy flux. This result points to a pencil beam pattern for the pulsars operating in a sub-critical regime. The estimated luminosity of \src\ during our \Nu\ observation is $L\sim2\times10^{37}$~\lum\ for the distance of 10 kpc (see below), whereas the critical luminosity for the pulsar with magnetic field $B\sim5\times10^{12}$\,G is expected to be around $3\times 10^{37}$ \lum\ \citep{2015MNRAS.447.1847M}. Thus one may conclude that \src\ was observed very close to the critical luminosity, but still in the sub-critical regime.

\subsection{Origin of the IR companion}

The study of optical catalogs and observational data showed the absence of any object in the localization region of the X-ray source. The upper limit on the observed magnitudes in filters $g, r, i$,  according to the Pan-STARRS instrumental filters, is around $23.1$.

  \begin{figure}
    \begin{center}
      \includegraphics[width=\columnwidth]{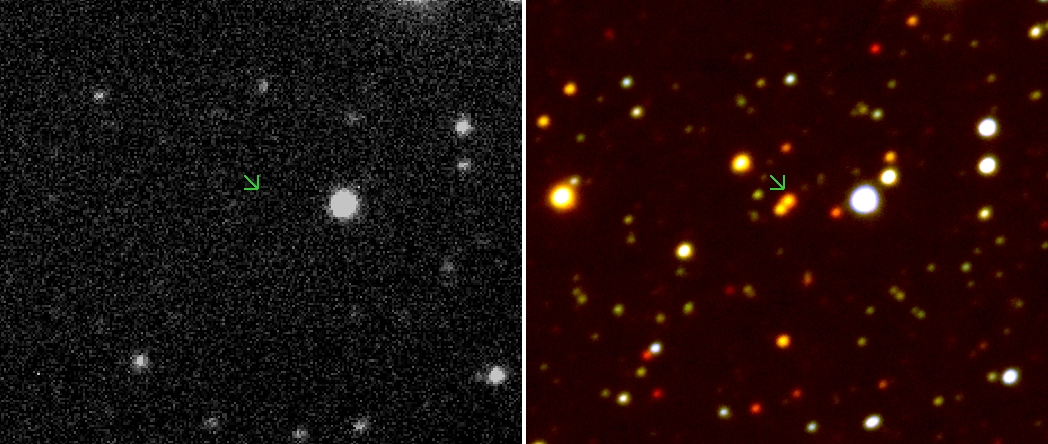}
    \end{center}
    \caption{Left: StanCam i-band image with exposure of 120 s. Right: NOTCam $JHK_s$ color coded image, each 540 s. Green arrow shows the IR-companion of \src. Both have N up, E left, and show the same region. }
    \label{fig:ijhk_not}
  \end{figure}

Inspecting the UKIDSS catalogue\footnote{http://wsa.roe.ac.uk/} we found that position of \src\ determined from the {\it Chandra} data is compatible with a faint infrared star, which in turn is apparently blended with another one (Fig.\,\ref{fig:ukinot}a). The blue circle illustrates an uncertainty of the source X-ray position. Based on the UKIDSS data it is impossible to separate fluxes from these stars to measure correct magnitudes of the counterpart in near-IR bands. Therefore to measure characteristics of the counterpart of \src\ we performed observations with the NOT telescope in several filters ($BVRiJHK_s$). The image of the sky field around \src, obtained with NOTCam in the $K_s$ filter is shown in Fig.~\ref{fig:ukinot}b. It is clearly seen, that the two objects are perfectly separated, by a distance of $0\farcs7$, and thus the only infrared companion of the source in the NOTCam data is a star with the coordinates of
RA=$18^{\rmn{h}}58^{\rmn{m}}43\fs63$, Dec=$3^\circ26\arcmin05\farcs2$ (J2000). 

The StanCam $BVRi$ images obtained at the NOT in good seeing show no detection at the location, see the $i$-band image compared to the $JHK_s$ color image in Fig.~\ref{fig:ijhk_not}. 
Photometric analysis of the NOT data using the PSF-photometry (DAOPHOT II) allowed us to determine magnitudes of this star as $J=18.559\pm0.037$, $H=15.291 \pm 0.035$, $K=13.520\pm0.027$. Note, that for calibrating instrumental magnitudes $JHK_s$ and converting them into $JHK$ ones we used the GPS/UKIDSS catalog as reference. 

\begin{figure}
\begin{center} 
\includegraphics[width=0.8\columnwidth, angle=270]{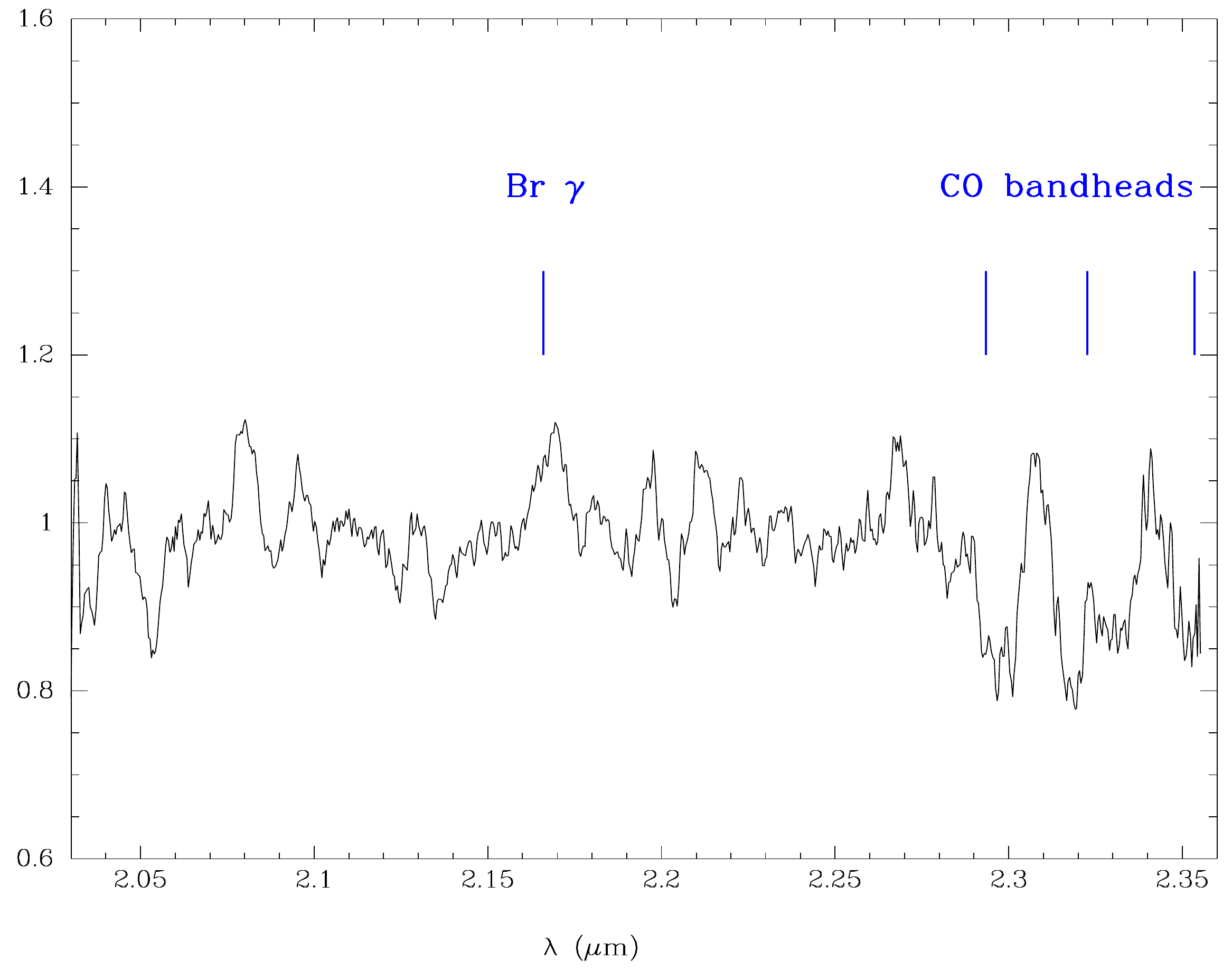}
\end{center}
\caption{$K$-band spectrum of the IR-counterpart of \src. The locations of the CO 2-0, 3-1, and 4-2 band heads at 2.29, 2.32 and 2.35 $\mu$m are indicated. }
\label{fig:ksspec} 
\end{figure}

The $K$-band spectrum obtained with NOTCam resulted in a very low S/N ratio of around 7 due to less than optimal conditions. Smoothing the spectrum over 17 pixels, however, and thereby lowering the resolving power from 2100 to 120, improved the S/N to 20, and revealed tentative detection of the CO 2-0, 3-1, and 4-2 bandheads in absorption at 2.29, 2.32 and 2.35 microns, respectively, of which the latter is cut halfway by the instrument sensitivity, see Fig.~\ref{fig:ksspec}. The RMS in the smoothed spectrum is of the order of 5\% and the depth of the features around 15\%, giving a 3$\sigma$ detection. As shown in the $K$-band spectral catalog of \citet{1997ApJS..111..445W} the presence of these CO bands in absorption strongly suggests a late spectral type of luminosity class I or III. 

The EW estimated for the 2-0 bandhead at 2.297 $\mu$m is 24 \AA\ which would point to a late M-type giant or early M-type supergiant according to the relation between spectral type and CO equivalent widths in Figure 2 in  \citet{2007ApJ...671..781D}. The 3-1 bandhead at 2.323 $\mu$m is even broader but also more contaminated with noise. We believe it is fairly safe to deduce that the object is a late type, most likely an M-type giant or supergiant. The rest of the spectrum remains featureless, although there is a small bump below detection levels at the position of Br-$\gamma$.

Some conclusions can also be made about the nature of the studied star based on the photometric results. Taking intrinsic colors $(H-K)_0$ of different classes of stars from \cite{2014AcA....64..261W,2015AN....336..159W}, we compare them with the $(H-K)$ color of the counterpart determined from the NOT observations. From a simple relation $E(H-K)=(H-K)-(H-K)_0$ we can find an extinction correction for each class of stars to correspond to the measured color of the source. 
Assuming a standard extinction law \citep{1989ApJ...345..245C} expected in this sky region, we can transform each $E(H-K)$ into $A_{K}$. At the same time, comparing absolute magnitudes in the $K$-band of the same stars \citep{2000MNRAS.319..771W,2006MNRAS.371..185W,2007MNRAS.374.1549W} with the measured magnitude of the source in the $K$-band (taking into account $A_{K}$), we can estimate a probable distance to each class of stars from a relation $5-5\log_{10}D=M_{\rm K, abs} - K_{\rm NOT} + A_{K}$.

\begin{figure}
\begin{center} 
\includegraphics[width=0.9\columnwidth]{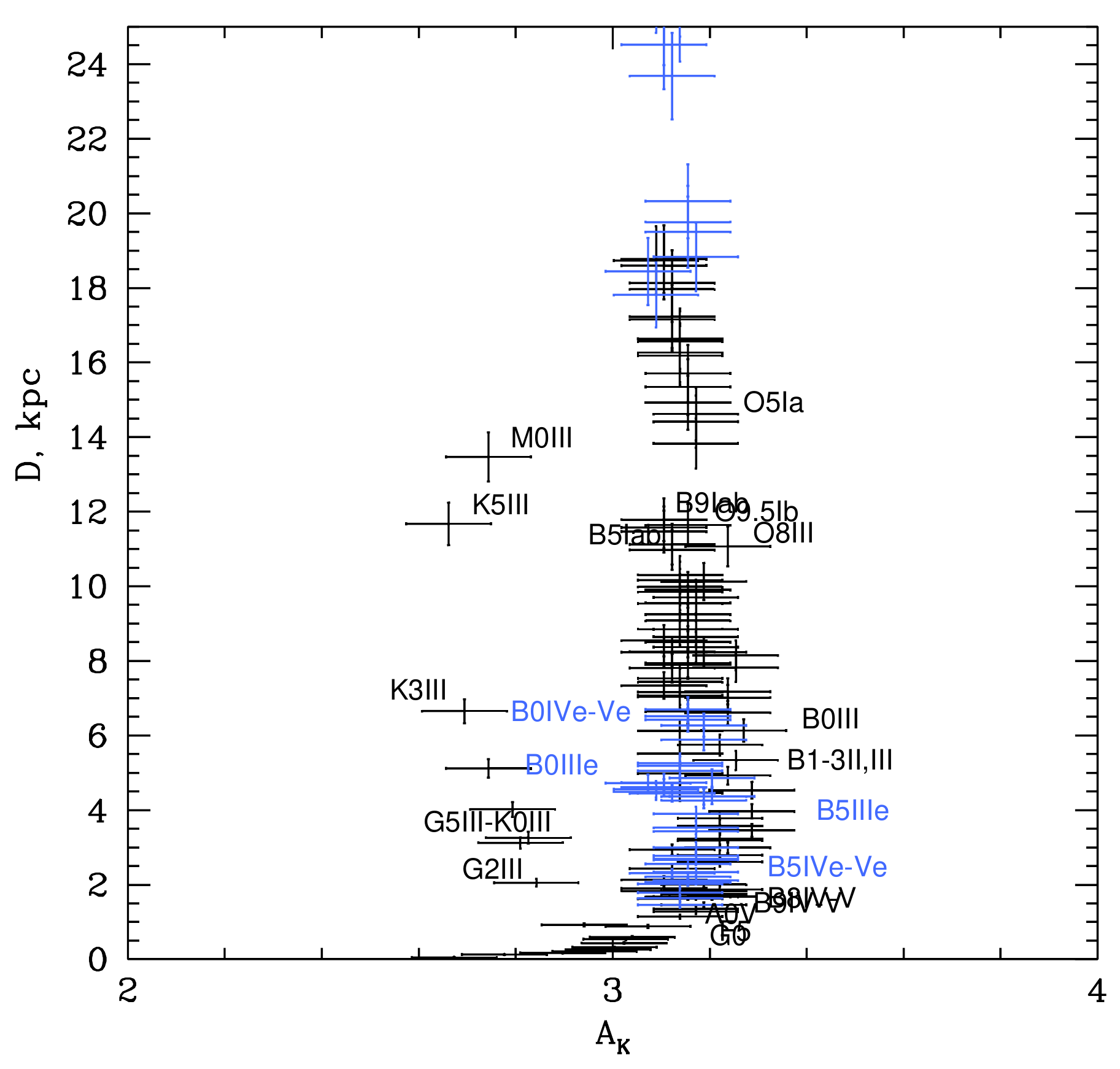}
\end{center}
\caption{Distance-absorption diagram showing how far a star of a specific class should be located if it is a counterpart of \src\ and the appropriate magnitude of the absorption towards such star. A subclass of Be-stars is marked in blue. 
}
\label{fig:irtype} 
\end{figure}

The results of such estimations for different classes of stars are presented in Fig.~\ref{fig:irtype}. From this diagram it is not possible to unambiguously determine the class of the companion, but we can get some restrictions on the extinction magnitude towards the source. In particular, for OB-giants or supergiants the extinction $A_{K}$ toward \src\ should be around $\sim(3.1-3.4)$, for red giant stars $A_{K}\simeq(2.6-3.0)$, and for main sequence stars $A_{K}$ is somewhere between these values. In particular, if we assume that the optical counterpart of \src\ is a giant star it should be located at $\sim$4--14\,kpc from the Sun. Note, that above extinction converted into the hydrogen column density $N_{\rm H}$ via standard relations  \citep{1995A&A...293..889P,2009MNRAS.400.2050G} agree generally with results obtained from the source X-ray spectrum in the low state. 

A significant absorption towards the source can easily explain its non-detection in optical filters at the NOT telescope and with {\it Gaia}. The extinction $A_{K}\simeq2.8$ corresponds to $A_{i}\simeq16$ for the standard law. Thus an expected magnitude of the source in the $i$-filter should be $\sim$30.

\begin{figure}
\begin{center} 
\includegraphics[width=0.9\columnwidth, trim={1.5cm 8.5cm 1cm 4cm},clip]{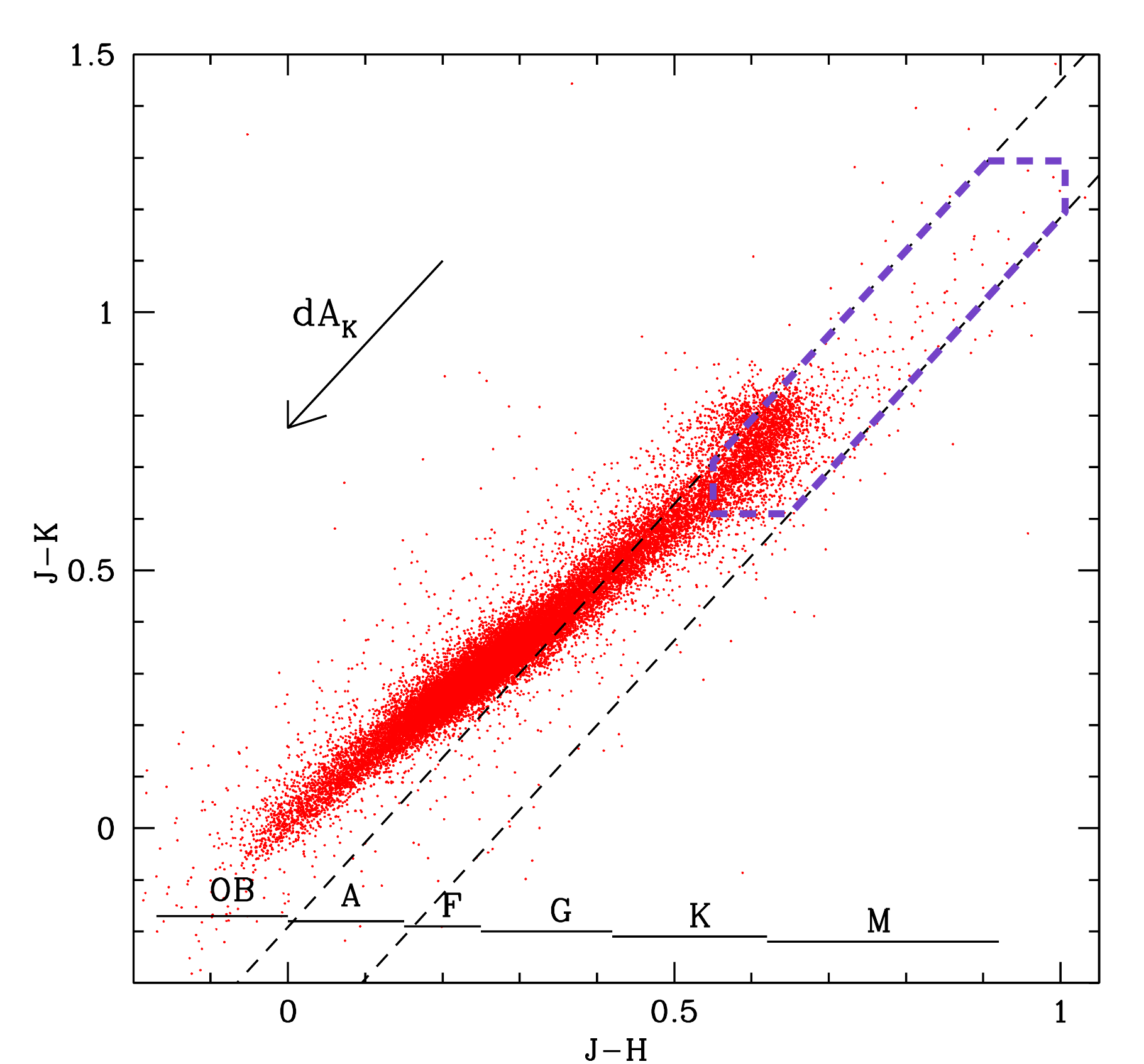}
\end{center}
\caption{Color-color diagram for unreddened stars (2MASS+Hiparcos) of various spectral types (marked by the strips with the letters of spectral types). The inclination of an arrow corresponds to standard extinction law  \citep{1989ApJ...345..245C}. Dashed lines track the variation of colors of counterpart of \src\ due to variation of the absorption correction. Violet polygon marks the most probable class of stars to be a real counterpart of the source.}
\label{fig:colors} 
\end{figure}

At the next step we compare the observed colors $(J-K)$ and $(J-H)$ of the counterpart with intrinsic ones of different classes of stars in a similar way as described by \citet{2012AstL...38..629K}. The red dots in Fig.\,\ref{fig:colors} indicate colors of the different type stars in the nearest 100 pc from the Sun taken from the 2MASS catalogue and {\it Hipparcos} observatory data \citep{2007A&A...474..653V}. The interstellar extinction for such nearby stars in the IR-filters is negligibly small, so we obtain the desired set of unabsorbed reference stars of different classes. The strips at the bottom indicate the
regions of the diagram corresponding to a particular
type of stars (marked by the letters, see, e.g., \citealt{2014AcA....64..261W,2015AN....336..159W}).

\begin{figure}
\includegraphics[width=0.75\columnwidth, angle =-90]{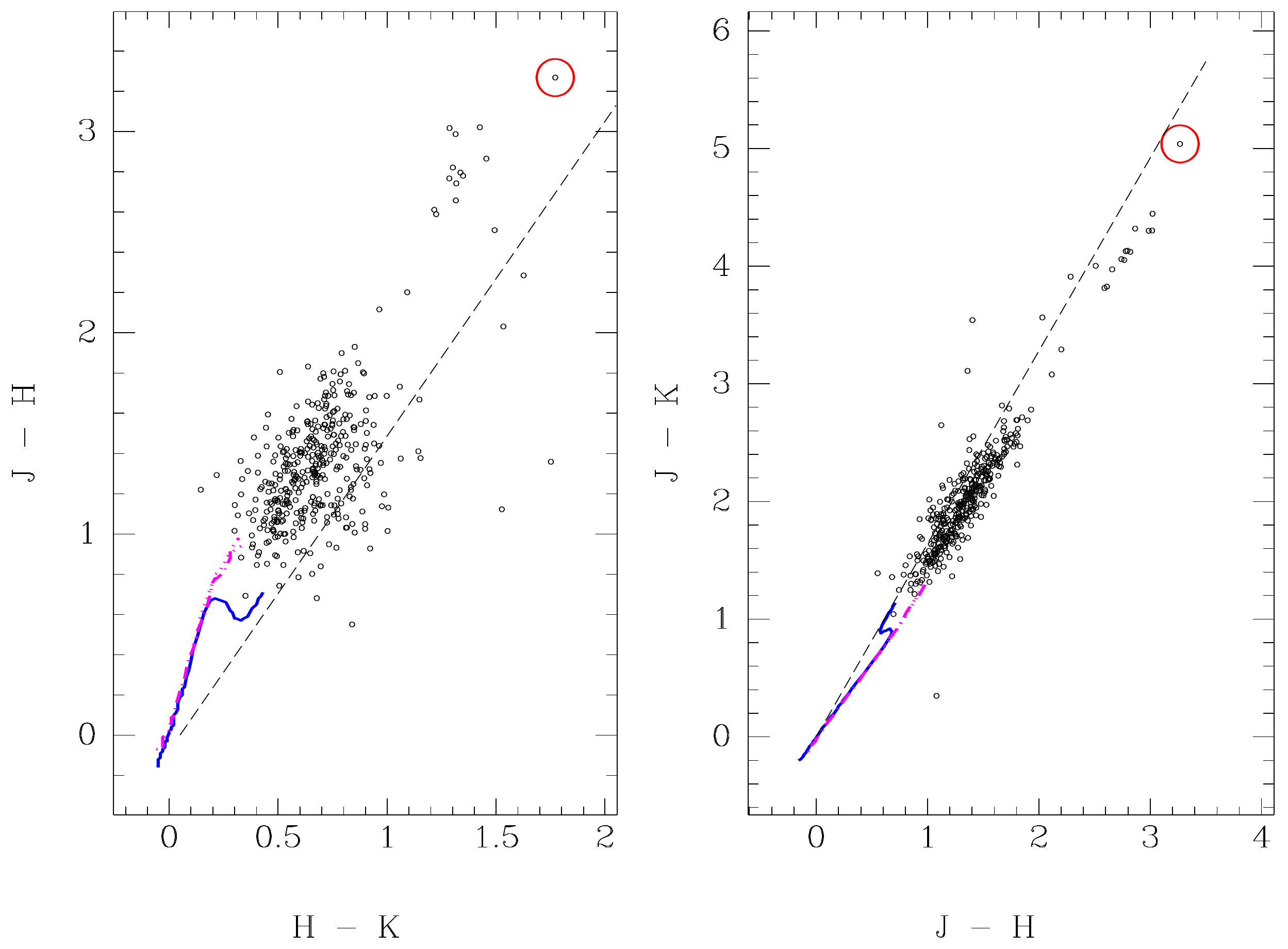}
\caption{Near-IR colour-colour diagrams of the 80$\arcsec$ region observed in $JHK$.
The target is encircled (red). The loci of main-sequence stars are shown as continuous blue curves, while those of giants and supergiants are shown as dotted and dot-dashed pink curves (they overlap). The extinction vector of an A0 star is shown as a dashed line.
}
\label{fig:XTEnir} 
\end{figure}

The position of the IR counterpart of \src\ ($J-H\simeq3.27$ and $J-K \simeq5.04$) is far beyond this diagram due to a significant interstellar extinction (see Fig.~\ref{fig:XTEnir}). But, correcting the source's magnitudes for different extinctions, we can shift its colors downward and to the left toward unreddened stars. The direction of this motion is indicated by the dashed straight lines whose slope corresponds to the ratio $E(J - K)/E(J - H)$ under the assumption of a standard extinction law \citep{1989ApJ...345..245C}.

The separation between the lines corresponds to the uncertainty of the measured colors of the source. Only the stars lying within the dashed lines can be the source’s companions. We see from the figure that only the red giants of late K or M classes are eligible to be a real companion of \src. At the same time, early classes of giants or supergiants can be excluded according to this diagram. 

As was previously shown, if the counterpart of \src\ is indeed a red giant star, the magnitude of the absorption towards the source is $A_{K}\simeq(2.6-3.0)$. Using these values for the extinction correction of the observed colors, we can additionally restrict the class of possible source companions (a violet dashed polygon in Fig. \ref{fig:colors}).

Summarizing all of the above, we can conclude that the most likely optical counterpart of the source is a K-M red giant located at a distance of 7--14 kpc. It is worth noting that such a significant distance to the system is supported by the X-ray timing analysis performed by Malacaria et al. (in press), whose estimate is $d\sim10.9$~kpc.



The same conclusion can be reached using a slightly different approach to the analysis and interpretation of photometric measurements. The $JHK$ photometry of point sources in the 80$\arcsec \times 80\arcsec$ FOV of the NOTCam HR-camera in the $J-H/H-K$ and $J-K/J-H$ color-color diagrams are shown in Fig.~\ref{fig:XTEnir}. In both diagrams the intrinsic colors of main-sequence stars are shown as a continuous blue curve, while those of giants (dotted pink curve) and super-giants (dot-dashed pink curve) taken from  \citet{1983A&A...128...84K}. Giants and supergiants overlap but both clearly separate from the main-sequence stars at late spectral types. The reddening vector is plotted for an A0 star as a dashed line based on a standard interstellar extinction law \citep{1989ApJ...345..245C}. The target is the encircled dot, one of the reddest objects detected in all three filters. We see that the target seems to follow the trace of a highly reddened late type giant or supergiant, and its location does not fit an early type reddened star. The photometry clearly supports the spectroscopic suggestion of a K or M giant or supergiant. The intrinsic $H-K$ color is in the small range from 0.12 to 0.35 mag from K0 to M6 stars according to \citet{1983A&A...128...84K} and \citet{2014AcA....64..261W}. The two extremes give $E(H-K)$ = $(H-K)_{\rm obs} - (H-K)_{\rm int}$ from 1.42 to 1.65 mag, which translates to extinction estimates of $A_{\rm K}$ in the range from 2.5 to 2.9 mag. This gives a dereddened $K$-band magnitude in the range from 10.6 to 11.0 mag. The different spectral types have different absolute $K$-band magnitudes, resulting in a range of distances from 7--14 kpc for K-M giants, while for supergiants the distance is even larger.

\subsection{Previously proposed companions}

At least two different stars were earlier considered as a possible companion of \src\ \citep{2005A&A...440..637R,2020ApJ...896...90M}. The first star was discussed above and is the only one in the nearby vicinity of \src\ (see Fig.\,\ref{fig:chima}, marked by a cross), whose optical spectrum shows H$\alpha$ emission. \cite{2005A&A...440..637R} suggested this object to be a Be star and a possible counterpart of \src. It is important to note that coordinates of this star reported by \citet{2005A&A...440..637R} are approximately 6\arcsec\ away from another X-ray object, registered by \ch\ inside the {\it INTEGRAL}/IBIS error circle (marked with ``2'' in Fig.\,\ref{fig:chima}, coordinates 
RA=$18^{\rmn{h}}58^{\rmn{m}}35\fs62$, Dec=$3^\circ26\arcmin10\farcs5$; 
\citealt{2014yCat....1.2023S}). We checked Pan-STARRS and {\it Gaia} data and found that there is only one optical star coinciding exactly with the \ch\ source ``2''. Its magnitudes in optical filters are similar to the ones reported by \citet{2005A&A...440..637R}. These facts allow us to suggest that this star and the star reported by \citet{2005A&A...440..637R} are the same object. The {\it Gaia} data indicate that it is located at a distance of $\simeq200$\,pc \citep{2018AJ....156...58B} and has an effective temperature of $T_{\rm eff}\simeq4000$\,K \citep{2018A&A...616A...1G}. These measurements as well as the registration of the H$\alpha$ emission line by \citet{2005A&A...440..637R}  indicate that it can be a nearby cataclysmic variable. 

Another optical companion of \src\ was considered by \citet{2020ApJ...896...90M} from the {\it Gaia} catalogue as the closest star to the nominal source position known at that moment (marked by ``X'' in Fig.\,\ref{fig:ukinot}). 
Based on the above analysis this hypothesis now can be firmly ruled out as well.
 
\section{Conclusion}

In this work we present the results of the spectral and temporal analysis of a poorly studied XRP \src\ performed in a broad range of energies and mass accretion rates. The spectrum of the source obtained with the \Nu\ observatory during an outburst in 2019 revealed the presence of a cyclotron absorption line in the energy spectrum at $E_{\rm cyc}\simeq48$ keV, that allowed us to estimate the NS magnetic field strength to be $5.2\times10^{12}$~G. 
The spectral properties of \src\ observed by the \xm\ and \ch\ observatories in the quiescent state point to ongoing accretion in this state, which we interpreted as accretion from the cold (low-ionization) accretion disk. 

\ch\ data allowed us to obtain the precise localisation of \src\ for the first time. Observations at the NOT telescope 
revealed only one potential near-IR companion of the pulsar. Spectral properties of the counterpart point to the red giant star located at 7--14 kpc suggesting that the system is likely a symbiotic binary hosting an XRP rather than Be X-ray binary as previously proposed. This distance agrees well with estimates obtained from the timing properties of the pulsar (Malacaria et al., in press).

\acknowledgements{
This work was supported by the grant 19-12-00423 of the Russian Science Foundation. 
Studies partially based on observations made with the Nordic Optical Telescope, owned in collaboration by the University of Turku and Aarhus University, and operated jointly by Aarhus University, the University of Turku and the University of Oslo, representing Denmark, Finland and Norway, the University of Iceland and Stockholm University at the Observatorio del Roque de los Muchachos, La Palma, Spain, of the Instituto de Astrofisica de Canarias.}


\bibliography{allbib}
\bibliographystyle{aasjournal}
\end{document}